\documentclass[osajnl,twocolumn,showpacs,superscriptaddress,10pt]{revtex4-1} 
\usepackage{amsmath,amssymb,graphicx}
\begin{document}

\title{Electromagnetic energy within single-resonance chiral metamaterial spheres}

\author{\firstname{Tiago}  J. \surname{Arruda}}
\email{tiagoarruda@pg.ffclrp.usp.br}
\affiliation{Faculdade de Filosofia,~Ci\^encias e Letras de Ribeir\~ao
Preto, Universidade de S\~ao Paulo, 14040-901, Ribeir\~ao
Preto-SP, Brazil}

\author{\firstname{Felipe}  A. \surname{Pinheiro}}
\affiliation{Instituto de F\'{i}sica, Universidade Federal do Rio de Janeiro, 21941-972, Rio de Janeiro-RJ, Brazil}

\author{\firstname{Alexandre} S. \surname{Martinez}}
\affiliation{Faculdade de Filosofia,~Ci\^encias e Letras de Ribeir\~ao
Preto, Universidade de S\~ao Paulo, 14040-901, Ribeir\~ao
Preto-SP, Brazil}
\affiliation{National Institute of Science and Technology in
Complex Systems}

\begin{abstract}

We derive an exact expression for the time-averaged electromagnetic energy inside a chiral dispersive sphere irradiated by a plane wave.
The dispersion relations correspond to a chiral metamaterial consisting of uncoupled single-resonance helical resonators.
Using a field decomposition scheme and a general expression for the electromagnetic energy density in bi-anisotropic media, we calculate the Lorenz-Mie solution for the internal fields in a medium that is simultaneously magnetic and chiral.
We also obtain an explicit analytical relation between the internal electromagnetic field and the absorption cross-section.
This result is applied to demonstrate that strong chirality leads to an off-resonance field enhancement within weakly absorbing spheres.

\end{abstract}


\maketitle


\section{Introduction}

Metamaterials are artificial structures with engineered electromagnetic (EM) response that may exhibit unusual properties such as negative refraction~\cite{Smith00}, image resolution beyond the diffraction limit~\cite{smith2004}, optical magnetism~\cite{enkrich2005,cai2007}, EM cloaking~\cite{pendry2006,leonhardt2006}, and slow light propagation~\cite{OPN}.
One possible route to achieve negative refraction requires that the real parts of the electric permittivity and magnetic permeability are simultaneously negative~\cite{Smith00,smith2004}.
This condition can be achieved with metamaterials composed of periodic structures containing both electric and magnetic resonators~\cite{Smith00}.
Another route for negative refraction involves single-resonance chiral metamaterials (CMMs), with neither electric permittivity nor magnetic permeability negative required~\cite{pendry2004}.
Although physically different, both routes explore resonant metamaterials, which are necessarily absorptive and dispersive.

The EM energy density in absorptive and dispersive media, which is crucial for many applications in photonics, was firstly derived for nonmagnetic materials~\cite{loudon} and then generalized for composite magnetic media~\cite{ruppin2002}.
For the particular case of wire-split ring resonator metamaterials, there basically exist two methods to calculate the EM energy density: the equivalent circuit~\cite{tretyakov} and the electrodynamic~\cite{boardman2006} approaches.
These methods were shown to be non-equivalent and the apparent inconsistency between them has been solved in~\cite{luan2009}.
Specially, this latter result has been recently used to calculate the EM energy within a coated magnetic sphere~\cite{tiago-joa}, an important geometry for applications in plasmonics.
The methodology developed in~\cite{luan2009} has been also employed to derive the EM energy density stored in single-resonance CMMs composed of uncoupled helical resonators~\cite{luan2011}.
To the best of our knowledge, the case of the EM energy within a dispersive sphere which is simultaneously chiral and magnetic, such as the CMM composites, has not been treated so far.

The aim of this paper is to derive an exact expression for the time-averaged EM energy inside a CMM sphere and its connection to the absorption cross-section.
Based on the Bohren's decomposition of the EM field for optically active spheres~\cite{bohren-active-sphere}, we obtain the Lorenz-Mie solution for the internal fields in the magnetic case~\cite{prl1,prl2}.
We investigate the behavior of internal resonances inside the sphere (near-field) and their corresponding resonances in the extinction efficiencies (far-field).
{In particular}, we derive an explicit expression for the absorption efficiency in terms of the internal EM fields that reveals that strong chirality leads to an off-resonance field enhancement within weakly absorbing spheres.

This paper is organized as follows.
In Sec.~\ref{theory}, we formulate the problem of EM scattering by spherical CMMs and derive the main analytical results of the paper.
In Sec.~\ref{numerical}, we present the numerical results, whereas Sec.~\ref{conclusions} is devoted to the conclusions.

\section{Basic theory}
\label{theory}

Let us consider a plane and monochromatic EM wave $[\mathbf{E}(\mathbf{r}),\mathbf{H}(\mathbf{r})]^{ T}\exp({-\imath\omega t})$, where $\omega$ is the angular frequency and $^T$ is the transpose operator.
This wave is incident to a linear, spatially homogeneous and isotropic CMM sphere, with radius $a$ and dispersive optical properties $[\epsilon_1(\omega),\mu_1(\omega),\kappa(\omega)]$,
where  $\epsilon_1$ and $\mu_1$ are the electric permittivity and magnetic permeability, respectively, and $\kappa$ is the dimensionless chirality parameter.
The surrounding medium is assumed to be the vacuum $(\epsilon_0,\mu_0)$ and the constitutive relations concerning to the sphere material are \cite{luan2011}
\begin{eqnarray}
    \mathbf{D}_1&=&\epsilon_1(\omega)\mathbf{E}_1+\imath\frac{\kappa(\omega)}{c}\mathbf{H}_1\ ,\label{const1}\\
    \mathbf{B}_1&=&\mu_1(\omega)\mathbf{H}_1-\imath\frac{\kappa(\omega)}{c}\mathbf{E}_1\
    ,\label{const2}
\end{eqnarray}
where $c=1/({\epsilon_0\mu_0})^{1/2}$ is the speed of light in vacuum and $\imath^2=-1$.
The refractive indices associated with the right-circularly polarized (RCP,+) and left-circularly polarized (LCP,$-$) waves are $m_1^{(\pm)}=({\epsilon_1/\epsilon_0})^{1/2}({\mu_1/\mu_0})^{1/2}\pm\kappa$.
Hence, for a real positive $\kappa$, it is possible to obtain negative refraction for LCP waves even if the real parts of $\epsilon_1$ and $\mu_1$ are positive~\cite{pendry2004}.

The constitutive relations in Eqs.~(\ref{const1}) and (\ref{const2}) for time-harmonic fields are based on the symmetrized Condon relations as discussed by Silverman~\cite{condon,silverman}.
Following~\cite{luan2011}, the constitutive parameters of a single-resonance chiral metamaterial are~\cite{zhao}:
\begin{eqnarray}
    \epsilon_1(\omega)&=&\epsilon_0\left[1-\frac{\omega_{\rm
    p}^2}{\omega^2-\omega_0^2+\imath\Gamma\omega}\right]\ ,\label{epsilon}\\
    \mu_1(\omega)&=&\mu_0\left[1-\frac{F\omega^2}{\omega^2-\omega_0^2+\imath\Gamma\omega}\right]\
    ,\label{mu}\\
    \kappa(\omega)&=&\frac{A\omega}{\omega^2-\omega_0^2+\imath\Gamma\omega}\
    ,\label{kappa}
\end{eqnarray}
where $\omega_{\rm p}$ and $\omega_0$ are the plasma and resonance frequencies of the resonators, $F$ is a dimensionless filling factor in one unit cell ($0<F<1$), $\Gamma$ is a dissipative coefficient and $A= \pm\omega_{\rm p}{F}^{1/2}$ \cite{luan2011,zhao}.
Both $\mu_1$ and $\kappa$ depend on the filling factor $F$; in particular, $|\kappa|^2\propto{F}$.
To guarantee the validity of these constitutive parameters, we assume that the relevant wavelength within the sphere is much longer than the lattice constant associated with one unit cell of the uncoupled helical resonators \cite{zhao}.

\subsection{Energy density in single-resonance chiral metamaterial}

For a harmonic wave in a medium $[\epsilon_1(\omega),\mu_1(\omega),\kappa(\omega)]$, described by Eqs.~(\ref{const1})--(\ref{kappa}), the time-averaged EM energy density is
\begin{eqnarray}
\langle u\rangle_t= \frac{1}{4}\left[\epsilon_{\rm eff}|\mathbf{E}_1|^2+\mu_{\rm eff}|\mathbf{H}_1|^2+\frac{2}{c}{\rm Re}\left(\kappa_{\rm eff}\mathbf{E}_1^*\cdot\mathbf{H}_1\right)\right]\ ,\label{density}
\end{eqnarray}
where the effective energy coefficients are given by \cite{luan2009,luan2011}
\begin{eqnarray}
    \epsilon_{\rm eff}(\omega) &=&\epsilon_0\left[1+\frac{\omega_{\rm
    p}^2\left(\omega_0^2+\omega^2\right)}{\left(\omega_0^2-\omega^2\right)^2+\Gamma^2\omega^2}\right]\
    ,\label{epsilon-eff}\\
    \mu_{\rm
    eff}(\omega)&=&\mu_0\left[1+\frac{F\omega^2\left(3\omega_0^2-\omega^2\right)}{\left(\omega_0^2-\omega^2\right)^2+\Gamma^2\omega^2}\right]\
    ,\label{mu-eff}\\
    \kappa_{\rm
    eff}(\omega)&=&-\frac{A\omega\left(\Gamma\omega+2\imath\omega_0^2\right)}{\left(\omega_0^2-\omega^2\right)^2+\Gamma^2\omega^2}\
    .\label{kappa-eff}
\end{eqnarray}

Under the lossless assumption $(\Gamma=0)$, these coefficients assume the well-known form \cite{serdyukov}: $\epsilon_{\rm eff}=\partial(\omega\epsilon_1)/\partial\omega>0$, $\mu_{\rm eff}=\partial(\omega\mu_1)/\partial\omega>0$ and $\kappa_{\rm eff}=\imath\partial(\omega\kappa)/\partial\omega$, with $\epsilon_{\rm eff}\mu_{\rm eff}>\kappa_{\rm eff}^2$.
Therefore, the time-averaged EM energy $\langle W\rangle_t$ within the metamaterial sphere is obtained from the integration of Eq.~(\ref{density}) with respect to the volume of the particle:
\begin{eqnarray}
    \langle W\rangle_t = \int_0^{2\pi}{\rm d}\phi\int_{-1}^1{\rm d}(\cos\theta)\int_0^a{\rm d}r\ r^2 \; \langle u\rangle_t \ ,
\end{eqnarray}
where $r$ is the radial variable, and $\theta$ and $\phi$ are the polar and azimuthal angle variables in a spherical coordinates system centered in the sphere, respectively, and $\langle u\rangle_t=\langle u\rangle_t(r,\cos\theta,\phi)$ is given by Eq.~(\ref{density}).
We emphasize that Eqs.~(\ref{epsilon-eff})--(\ref{kappa-eff}) are not general, since they are derived for a special kind of CMM medium with constitutive relations and parameters given in Eqs.~(\ref{const1})--(\ref{kappa}).

\subsection{Electromagnetic scattering by an optically active sphere}

The exact solution for the EM wave scattering by an optically active sphere is developed in~\cite{bohren-active-sphere}.
Here, we generalize this solution to dispersive media that are both chiral and magnetic, and explicitly calculate the multipole moments associated with the internal EM field.
Following~\cite{bohren-active-sphere}, the sphere constitutive equations are
\begin{eqnarray}
    \mathbf{D}_1&=&\epsilon_1^{\rm(DBF)}\left(\mathbf{E}_1 + \beta\boldsymbol{\nabla}\times\mathbf{E}_1\right)\;,\label{const1-quiral}\\
    \mathbf{B}_1&=&\mu_1^{\rm(DBF)}\left(\mathbf{H}_1 +
    \beta\boldsymbol{\nabla}\times\mathbf{H}_1\right)\;,\label{const2-quiral}
\end{eqnarray}
which are known as Drude-Born-Fedorov (DBF) relations \cite{lakhtakia,lekner}.
The quantities $\epsilon_1^{\rm(DBF)}$, $\mu_1^{\rm(DBF)}$ and $\beta$ are phenomenological coefficients that describe the optical activity.
Equations~(\ref{const1-quiral}) and (\ref{const2-quiral}) are different from the corresponding constitutive Eqs.~(\ref{const1}) and (\ref{const2}).
However, assuming the time-harmonic dependence $e^{-\imath\omega t}$, the curl macroscopic Maxwell's equations $\boldsymbol{\nabla}\times\mathbf{E}_1=\imath\omega\mathbf{B}_1$ and $\boldsymbol{\nabla}\times\mathbf{H}_1=-\imath\omega\mathbf{D}_1$ hold for both set of constitutive relations.
Setting $\alpha=\beta\omega$ and $\gamma=\kappa/c$, one can easily demonstrate that the Condon and DBF constitutive relations are equivalent by the substitutions \cite{lakhtakia,lekner}
\begin{eqnarray}
    \epsilon_1^{\rm(DBF)}&=&\epsilon_1-\frac{\gamma^2}{\mu_1}\ ,\label{epsilon-dbf}\\
    \mu_1^{\rm(DBF)}&=&\mu_1-\frac{\gamma^2}{\epsilon_1}\ ,\label{mu-dbf}\\
    \alpha&=&\frac{\gamma}{\epsilon_1\mu_1-\gamma^2}\
    ,\label{alpha-dbf}
\end{eqnarray}
with inverse relations \cite{lakhtakia}: $\epsilon_1={\epsilon_1^{\rm(DBF)}}/\zeta$, $\mu_1={\mu_1^{\rm(DBF)}}/\zeta$ and $\gamma={\alpha\epsilon_1^{\rm(DBF)}\mu_1^{\rm(DBF)}}/\zeta$, where $\zeta={1-\alpha^2\epsilon_1^{\rm(DBF)}\mu_1^{\rm(DBF)}}$.

The relations above between the constitutive parameters in DBF approach and Eqs.~(\ref{const1}) and (\ref{const2}) allow us to write the Lorenz-Mie quantities in the system of parameters $[\epsilon_1^{\rm(DBF)},\mu_1^{\rm(DBF)},\beta]$ and calculate the time-averaged EM energy in the system of parameters $(\epsilon_1,\mu_1,\kappa)$.
According to~\cite{bohren,bohren-active-sphere}, we can relate these phenomenological coefficients to the complex refractive indices $m_{\rm L}$ and $m_{\rm R}$ through
\begin{eqnarray}
    \beta&=&\frac{1}{2}\left(\frac{1}{k_{\rm R}}-\frac{1}{k_{\rm L}}\right)\
    ,\\
    \omega\left[{\epsilon_1^{\rm(DBF)}\mu_1^{\rm(DBF)}}\right]^{1/2}&=&\left[\frac{1}{2}\left(\frac{1}{k_{\rm R}}+\frac{1}{k_{\rm L}}\right)\right]^{-1}\ ,
\end{eqnarray}
where $k_q=k m_q$ is the wave number in the active medium $[\epsilon_1^{\rm(DBF)},\mu_1^{\rm(DBF)},\beta]$, with $q$ is either ${\rm L}$ or ${\rm R}$ for refractive indices associated with left-circularly polarized (LCP,$-$) or right-circularly polarized (RCP,$+$) waves, respectively, and $k$ is the incident wave number.
Explicitly, one has $1/k_{\rm R}=1/k_1^{\rm (DBF)}+\beta$ and $1/k_{\rm L}=1/k_1^{\rm (DBF)}-\beta$, with $k_1^{\rm (DBF)}=\omega[{\epsilon_1^{\rm(DBF)}\mu_1^{\rm(DBF)}}]^{1/2}$.
Since the real parts of $\epsilon_1^{\rm(DBF)}$ and $\mu_1^{\rm(DBF)}$ can both assume negative values, it is convenient to write $k_1^{\rm (DBF)}=\omega[{\epsilon_1^{\rm(DBF)}}]^{1/2}[{\mu_1^{\rm(DBF)}}]^{1/2}$.

The macroscopic Maxwell's equations for harmonic waves in DBF approach are $({\nabla}^2+\mathcal{K}^2)(\mathbf{E}_1,\mathbf{H}_1)^{T}=(\mathbf{0},\mathbf{0})^T$, $\boldsymbol{\nabla}\times(\mathbf{E}_1,\mathbf{H}_1)^T=\mathcal{K}(\mathbf{E}_1,\mathbf{H}_1)^T$, $\boldsymbol{\nabla}\cdot(\mathbf{E}_1,\mathbf{H}_1)^T=(0,0)^T$, where \cite{bohren}
\begin{equation*}
\mathcal{K} =
    \dfrac{\imath\omega}{\zeta}\begin{pmatrix}-\imath\alpha\epsilon_1^{\rm(DBF)}\mu_1^{\rm(DBF)}&\mu_1^{\rm(DBF)}\\-\epsilon_1^{\rm(DBF)}&-\imath\alpha\epsilon_1^{\rm(DBF)}\mu_1^{\rm(DBF)}\end{pmatrix}\ .
\end{equation*}
To diagonalize $\mathcal{K}$ and simplify the system of equations, one can use a linear transformation of the EM field~\cite{bohren-active-sphere,bohren}: $(\mathbf{E}_1,\mathbf{H}_1)^T=\mathcal{A}(\mathbf{Q}_{\rm L},\mathbf{Q}_{\rm R})^T$, with
$\mathcal{A}^{-1}\mathcal{K}\mathcal{A}=\left(\begin{array}{cc}k_{\rm L}&0\\0&-k_{\rm R}\end{array}\right)$,
where $\mathcal{A}  = \left( \begin{array}{cc}1&\alpha_{\rm R}\\\alpha_{\rm L}&1\end{array} \right)$,
$\alpha_{\rm R}  =  -\imath[{{\mu_1^{\rm(DBF)}}/{\epsilon_1^{\rm(DBF)}}}]^{1/2}$ and $\alpha_{\rm L}= {-1}/{\alpha_{\rm R}}$.
The EM left-handed and right-handed waves $\mathbf{Q}_{\rm L}$ and $\mathbf{Q}_{\rm R}$, respectively, are the vector basis for the Bohren's decomposition scheme and independently satisfy the vector Helmholtz equation $(\nabla^2+k^2)\mathbf{Q}=\mathbf{0}$, with $\boldsymbol{\nabla}\times\mathbf{Q}=k\mathbf{Q}$ and $\boldsymbol{\nabla}\cdot\mathbf{Q}=0$, where $k=k_{\rm L}$ for $\mathbf{Q}=\mathbf{Q}_{\rm L}$ and $k=-k_{\rm R}$ for $\mathbf{Q}=\mathbf{Q}_{\rm R}$.
Solving these equations for the transformed fields $\mathbf{Q}$ and returning to the original internal EM field $(\mathbf{E}_1,\mathbf{H}_1)^T$, one obtains in the spherical coordinates system $(r,\theta,\phi)$ the electric field components:
\begin{eqnarray}
    E_{1r} &=& \sum_{n=1}^{\infty}E_n\sin\theta n(n+1)\pi_n\nonumber\\
    &&  \Bigg\{\frac{\psi_n(\rho_{\rm L})}{\rho_{\rm L}^2}\left[f_{{\rm o}n}\sin\phi+f_{{\rm e}n}\cos\phi\right]\nonumber\\
    &&-\alpha_{\rm R}\frac{\psi_n(\rho_{\rm R})}{\rho_{\rm R}^2}\left[g_{{\rm o}n}\sin\phi+g_{{\rm e}n}\cos\phi\right]\Bigg\}\ ,\label{Er}\\
    E_{1\theta}&=& \sum_{n=1}^{\infty}E_n\Bigg\{\frac{\cos\phi}{\rho_{\rm L}}\left[f_{{\rm o}n}\pi_n\psi_n(\rho_{\rm L}) +f_{{\rm e}n}\tau_n\psi_n'(\rho_{\rm L})\right] \nonumber \\
    &&+\frac{\sin\phi}{\rho_{\rm L}}\left[f_{{\rm o}n}\tau_n\psi_n'(\rho_{\rm L})-f_{{\rm e}n}\pi_n\psi_n(\rho_{\rm L})\right]\nonumber\\
    &&+ \alpha_{\rm R} \frac{\cos\phi}{\rho_{\rm R}} \left[g_{{\rm o}n}\pi_n\psi_n(\rho_{\rm R})-g_{{\rm e}n}\tau_n\psi_n'(\rho_{\rm R})\right]\nonumber\\
    && - \alpha_{\rm R} \frac{\sin\phi}{\rho_{\rm R}}\left[g_{{\rm o}n}\tau_n\psi_n'(\rho_{\rm R})+g_{{\rm e}n}\pi_n\psi_n(\rho_{\rm R})\right]\Bigg\}\ ,\\
    E_{1\phi} & =&\sum_{n=1}^{\infty}E_n\Bigg\{\frac{\cos\phi}{\rho_{\rm
    L}}\left[f_{{\rm o}n}\pi_n\psi_n'(\rho_{\rm L})-f_{{\rm
    e}n}\tau_n\psi_n(\rho_{\rm
    L})\right] \nonumber\\
    &&-\frac{\sin\phi}{\rho_{\rm L}}\left[f_{{\rm
    o}n}\tau_n\psi_n(\rho_{\rm L})+f_{{\rm
    e}n}\pi_n\psi_n'(\rho_{\rm L})\right]\nonumber
    \\ &&-
    \alpha_{\rm R}\frac{\cos\phi}{\rho_{\rm
    R}}\left[g_{{\rm
    o}n}\pi_n\psi_n'(\rho_{\rm R})+g_{{\rm
    e}n}\tau_n\psi_n(\rho_{\rm R})\right]\nonumber\\
    &&  - \alpha_{\rm R}\frac{\sin\phi}{\rho_{\rm
    R}} \left[g_{{\rm
    o}n}\tau_n\psi_n(\rho_{\rm R})-g_{{\rm
    e}n}\pi_n\psi_n'(\rho_{\rm R})\right]\Bigg\}\ ,\label{Ephi}
%
\end{eqnarray}
where $E_n=E_0\imath^n(2n+1)/[n(n+1)]$, $\rho_q=k_qr$, with $q$ is either ${\rm L}$ or R, and $E_0$ is the amplitude of the incident EM wave.
The corresponding magnetic field $\mathbf{H}_1$ is obtained from Eqs.~(\ref{Er})--(\ref{Ephi}) by replacing $(f_n,\alpha_{\rm R}g_n)$ with $(\alpha_{\rm L}f_n,g_n)$.
The radial functions $\psi_n(\rho)=\rho j_n(\rho)$ and $\xi_n(\rho)=\rho[j_n(\rho)+\imath y_n(\rho)]$ are the Riccati-Bessel and Riccati-Hankel functions, respectively, where $j_n$ and $y_n$ are the spherical Bessel and Neumann functions.
The angular functions are $\pi_n={P_n^1(\cos\theta)}/{\sin\theta}$ and $\tau_n={{\rm d}}{P_n^1(\cos\theta)}/{{\rm d}\theta}$, with $P_n^1(\cos\theta)$ being the associated Legendre function of first order.
Expansions of the incident and scattered EM fields in terms of vector spherical harmonics can be found in~\cite{bohren}.

Using Bohren's notation~\cite{bohren-active-sphere}, the Lorenz-Mie coefficients for the splitted internal fields are:
\begin{eqnarray}
    f_{{\rm o}n}&=&\frac{\imath m_{\rm L} W_n^{\rm(R)}}{W_n^{\rm(L)}V_n^{\rm(R)}+W_n^{\rm(R)}V_n^{\rm(L)}}\;,\label{fn-quiral}\\
    f_{{\rm e}n}&=&\frac{m_{\rm L} V_n^{\rm(R)}}{W_n^{\rm(L)}V_n^{\rm(R)}+W_n^{\rm(R)}V_n^{\rm(L)}}\;,\\
    g_{{\rm o}n}&=&\frac{- \imath m_{\rm R} W_n^{\rm(L)} \alpha_{\rm L} }{W_n^{\rm(L)}V_n^{\rm(R)}+W_n^{\rm(R)}V_n^{\rm(L)}}\;,\\
    g_{{\rm e}n}&=&\frac{ m_{\rm R}
    V_n^{\rm(L)} \alpha_{\rm L} }{W_n^{\rm(L)}V_n^{\rm(R)}+W_n^{\rm(R)}V_n^{\rm(L)}}\;,\label{gn-quiral}\\
    V_n^{ (q)}&=&\psi_n(m_{
    q}x)\xi_n'(x)-\widetilde{m}\xi_n(x)\psi_n'(m_qx)\ ,\\
    W_n^{ (q)}&=&\widetilde{m}\psi_n(m_{
    q}x)\xi_n'(x)-\xi_n(x)\psi_n'(m_qx)\ ,
\end{eqnarray}
where $x=ka$ is the size parameter and the effective impedance index $\widetilde{m}$ is
\begin{eqnarray*}
{\widetilde{m}}=\left[{\frac{\mu_0\epsilon_1^{\rm(DBF)}}{\epsilon_0\mu_1^{\rm(DBF)}}}\right]^{1/2}=\frac{\mu_0}{\mu_1^{\rm(DBF)}}\left[\frac{1}{2}\left(\frac{1}{m_{\rm
R}}+\frac{1}{m_{\rm L}}\right)\right]^{-1}\ ,
\end{eqnarray*}
with $m=\widetilde{m}\mu_1^{\rm(DBF)}/\mu_0$ being the effective refractive index associated with the chiral medium $[\epsilon_1^{\rm(DBF)},\mu_1^{\rm(DBF)},\beta]$.
In particular, if the medium is nonmagnetic [$\mu_1^{\rm(DBF)}=\mu_0$], one has $m=\widetilde{m}$~\cite{bohren}.
The internal coefficients $g_{{\rm o}n}$ and $g_{{\rm e}n}$ have dimensions of $\alpha_{\rm L}$ and, thereby, it is expected to appear in the dimensionless calculated quantities the products $\alpha_{\rm R}g_{{\rm o}n}$ and $\alpha_{\rm R}g_{{\rm e}n}$.

\subsection{Electromagnetic energy within chiral spheres}

To calculate the time-averaged EM energy within a spherical particle, we follow the same procedure of~\cite{bott,tiago-sphere,tiago-cylinder,tiago-joa}.
We define the partial contributions to the average EM energy $\langle W\rangle_t$, since this quantity can be written as a sum of the electric, magnetic and magnetoelectric coupling terms:
\begin{equation}
    \langle W\rangle_t = \langle W_E\rangle_t + \langle
    W_H\rangle_t + \langle W_{EH}\rangle_t\ ,\label{ener}
\end{equation}
where $\langle W_{E}\rangle_t = {\epsilon_{\rm eff}}\int_{\mathcal{V}}{\rm d}^3r |\mathbf{E}_1|^2/4$, $\langle W_{H}\rangle_t = {\mu_{\rm eff}} \int_{\mathcal{V}}{\rm d}^3r|\mathbf{H}_1|^2/4$ and $\langle W_{EH}\rangle_t =\int_{\mathcal{V}}{\rm d}^3r{\rm Re}(\sigma_{\rm eff}\mathbf{H}_1\cdot\mathbf{E}_1^*)/2c$, with the region of integration $\mathcal{V}$ being the volume of the chiral sphere.
For a sphere with radius $a$ having the same optical properties as the surrounding medium $(\epsilon_0,\mu_0)$, the average EM energy is  $\langle W_{0}\rangle_t = {2}\pi a^3\epsilon_0|E_0|^2/{3}$ \cite{bott,tiago-sphere}.

In particular, consider the radial and angular contributions to the electric energy: $\langle W_E\rangle_t=\langle W_{Er}\rangle_t + \langle W_{E\theta}+W_{E\phi}\rangle_t$.
Explicitly, the radial contribution is
\begin{equation}\begin{split}
    \langle W_{Er}\rangle_t&=\frac{\epsilon_{\rm eff}}{4}\int_0^{2\pi}{\rm d}\phi\int_{-1}^1{\rm d}(\cos\theta)\int_0^a{\rm d}r\
    r^2|{E}_{1r}|^2\\
    &=\frac{3}{4a^3}{\langle W_0\rangle_t}\frac{\epsilon_{\rm eff}}{\epsilon_0}\sum_{n=1}^{\infty}n(n+1)(2n+1)\\
    &\int_0^a{\rm d}r\Bigg\{\left(\left|f_{{\rm
    o}n}\right|^2+\left|f_{{\rm
    e}n}\right|^2\right)\frac{|j_n(\rho_{\rm
    L})|^2}{|k_{\rm L}|^2}\\
    & +\left|\alpha_{\rm R}\right|^2\left(\left|g_{{\rm
    o}n}\right|^2+\left|g_{{\rm e}n}\right|^2\right)\frac{|j_n(\rho_{\rm R})|^2}{|k_{\rm R}|^2}\\
    & -2{\rm
    Re}\Bigg[\alpha_{\rm R}^*\left(f_{{\rm
    o}n}g_{{\rm o}n}^*+f_{{\rm e}n}g_{{\rm
    e}n}^*\right)\frac{j_n(\rho_{\rm L})j_n(\rho_{\rm R}^*)}{k_{\rm L}k_{\rm R}^*}\Bigg]\Bigg\}\
    ,\label{Wer}
\end{split}\end{equation}
where we have used $({2n+1})\int_{-1}^{1} {\rm d}(\cos \theta) \pi_n   \pi_{n'}\sin^2 \theta  = {2n(n+1)}\delta_{n,n'}$ to simplify this expression~\cite{bohren,tiago-sphere}.
Analogously, the angular contribution to the electric energy is $\langle W_{E\theta}+W_{E\phi}\rangle_t={\epsilon_{\rm eff}}\int_{\mathcal{V}}{\rm d}^3r(|{E}_{1\theta}|^2+|{E}_{1\phi}|^2)/4$, which leads
\begin{equation}\begin{split}
     \langle W_{E\theta}&+W_{E\phi}\rangle_t=\frac{3}{4a^3}\langle W_0\rangle_t\frac{{\epsilon_{\rm eff}}}{\epsilon_0}\sum_{n=1}^{\infty}(2n+1)\\
    &\ \int_0^a{\rm
    d}r\Bigg\{\left(|f_{{\rm
    o}n}|^2+|f_{{\rm e}n}|^2\right)\frac{\left|\psi_n(\rho_{\rm
    L})\right|^2+\left|\psi_n'(\rho_{\rm
    L})\right|^2}{|k_{\rm L}|^2}\\
    &\ +\left|\alpha_{\rm R}\right|^2\left(|g_{{\rm o}n}|^2+|g_{{\rm
    e}n}|^2\right)\frac{\left|\psi_n(\rho_{\rm
    R})\right|^2+\left|\psi_n'(\rho_{\rm
    R})\right|^2}{|k_{\rm R}|^2}\\
    &\ -2{\rm Re}\Bigg[\alpha_{\rm R}^*\left(f_{{\rm o}n}g_{{\rm o}n}^*+f_{{\rm
    e}n}g_{{\rm e}n}^*\right)\\
    &\ \frac{\psi_n'(\rho_{\rm L})\psi_n'(\rho_{\rm
    R}^*)-\psi_n(\rho_{\rm L})\psi_n(\rho_{\rm
    R}^*)}{k_{\rm L}k_{\rm R}^*}\Bigg]\Bigg\}\ ,\label{Wea}
\end{split}\end{equation}
where we have used the following relations \cite{tiago-sphere}: $({2n+1})\int_{-1}^{1} {\rm d}(\cos \theta) (\pi_n   \pi_{n'} + \tau_n  \tau_{n'} ) =  {2n^2(n+1)^2}\delta_{n,n'}$
and $\int_{-1}^{1} {\rm d}(\cos \theta) (\pi_n \tau_{n'} + \tau_n    \pi_{n'} ) = 0$.

From recurrence relations involving spherical Bessel functions, $(2n+1)j_n(\rho)=\rho[j_{n-1}(\rho)+j_{n+1}(\rho)]$ and $(2n+1)j_n'(\rho)=nj_{n-1}(\rho)-(n+1)j_{n+1}(\rho)$ \cite{tiago-sphere}, we obtain
\begin{equation}\begin{split}
    &\frac{(2n+1)}{\rho_{\rm A}\rho_{\rm B}^*}\left[n(n+1)j_n(\rho_{\rm A})j_n(\rho_{\rm B}^*)+\psi_n'(\rho_{\rm
    A})\psi_n'(\rho_{\rm B}^*)\right]\\
    &=nj_{n+1}(\rho_{\rm
    A})j_{n+1}(\rho_{\rm B}^*)+(n+1)j_{n-1}(\rho_{\rm
    A})j_{n-1}(\rho_{\rm B}^*)\;, \label{bessel}
\end{split}\end{equation}
where A and B can be either L or R.
Note that the left-hand side of Eq.~(\ref{bessel}) appears when we calculate the average electric energy $\langle W_E\rangle_t$ as the sum of Eqs.~(\ref{Wer}) and (\ref{Wea}).
Indeed, the right-hand side of Eq.~(\ref{bessel}) is used to simplify the calculations, once we have the analytical solution for the integral \cite{watson,tiago-sphere}
\begin{equation}\begin{split}
    \int_{0}^{a}{\rm d}r\ r^2&j_n(\rho_{\rm A})j_n(\rho_{\rm B}^*)\\
    &=a^3\frac{\left[{y_{\rm B}^*j_n(y_{\rm A})j_n'(y_{\rm B}^*)-{y_{\rm A}j_n(y_{\rm B}^*)j_n'(y_{\rm A})}}\right]}{y_{\rm A}^2-y_{\rm B}^{*2}}\ ,\label{int-ativa1}
\end{split}\end{equation}
with $y_{\rm A}=m_{\rm A}x$, $y_{\rm B}=m_{\rm B}x$, and we assume $m_{\rm A}\not=\pm m_{\rm B}^*$.
For $m_{\rm A}=\pm m_{\rm B}^*$, the L'Hospital's rule provides~\cite{tiago-sphere}
\begin{equation}\begin{split}
  \lim_{m_{\rm A}\to\pm m_{\rm B}^*}&\int_{0}^{a}{\rm d}r\ r^2j_n(m_{\rm A}kr)j_n(m_{\rm B}^*kr)\\
  &=\frac{a^3\imath^{(-1\pm1)n}}{2}\left[j_n^2(y_{\rm A})-j_{n-1}(y_{\rm A})j_{n+1}(y_{\rm A})\right]\ .\label{int-ativa2}
\end{split}\end{equation}
Therefore, defining the dimensionless functions
\begin{eqnarray}
    \mathcal{I}_{n}^{\rm(AB)}&=&\frac{1}{a^3}\int_{0}^{a}{\rm
    d}r\ r^2j_n(m_{\rm A}kr)j_n(m_{\rm B}^*kr)\ ,\\
    \mathcal{F}_{n,\pm}^{\rm(AB)}&=&n\mathcal{I}_{n+1}^{\rm(AB)}+(n+1)\mathcal{I}_{n-1}^{\rm(AB)}\pm(2n+1)\mathcal{I}_n^{\rm(AB)}\
    ,\\
    \mathcal{S}^{(\pm)}&=&\sum_{n=1}^{\infty}\bigg\{\left(\left|f_{{\rm
    o}n}\right|^2+\left|f_{{\rm
    e}n}\right|^2\right)\mathcal{F}_{n,+}^{\rm(LL)}\nonumber\\
    &&  +\left|\alpha_{\rm R}\right|^2\left(\left|g_{{\rm
    o}n}\right|^2+\left|g_{{\rm
    e}n}\right|^2\right)\mathcal{F}_{n,+}^{\rm(RR)}\nonumber \\
    & & \pm2{\rm
    Re}\left[\alpha_{\rm R}^*\left(f_{{\rm
    o}n}g_{{\rm o}n}^*+f_{{\rm e}n}g_{{\rm
    e}n}^*\right)\mathcal{F}_{n,-}^{\rm(LR)}\right]\bigg\}\ ,\label{S}
\end{eqnarray}
and summing $\langle W_E\rangle_t=\langle W_{Er}\rangle_t+\langle W_{E\theta}+W_{E\phi}\rangle_t$, from Eqs.~(\ref{Wer}) and (\ref{Wea}), and using Eq.~(\ref{bessel}), we finally obtain
\begin{eqnarray}
    \frac{\langle W_{E}\rangle_t}{\langle W_0\rangle_t}&=&\frac{3}{4}\frac{\epsilon_{\rm eff}}{\epsilon_0}\mathcal{S}_{E}
    ,\label{We}
\end{eqnarray}
where $S_{E}=S^{(-)}$ is a sum associated with the volume integral of $|\mathbf{E}_1|^2$.
Analogously, from the calculations of $\langle W_H\rangle_t=\langle W_{Hr}\rangle_t+\langle W_{H\theta}+W_{H\phi}\rangle_t$ and $\langle W_{EH}\rangle_t=\langle W_{EHr}\rangle_t+\langle W_{EH\theta}+W_{EH\phi}\rangle_t$, we obtain the magnetic and magnetoelectric coupling terms:
\begin{eqnarray}
    \frac{\langle W_{H}\rangle_t}{\langle W_0\rangle_t}&=&\frac{3}{4}\left|\widetilde{m}\right|^2\frac{\mu_{\rm eff}}{\mu_0}\mathcal{S}_{H} ,\label{Wh}\\
    \frac{\langle W_{EH}\rangle_t}{\langle W_0\rangle_t}&=&\frac{3}{2}{\rm Im}\left(\widetilde{m}\kappa_{\rm eff}\mathcal{S}_{EH}\right)\
    ,
    \label{Weh}
\end{eqnarray}
where $\mathcal{S}_{H}=S^{(+)}$ and
\begin{eqnarray}
\mathcal{S}_{EH}&=&\sum_{n=1}^{\infty}\bigg\{\left(\left|f_{{\rm o}n}\right|^2+\left|f_{{\rm  o}n}\right|^2\right)\mathcal{F}_{n,+}^{\rm(LL)}\nonumber\\
&&-\left|\alpha_{\rm R}\right|^2\left(\left|g_{{\rm o}n}\right|^2+\left|g_{{\rm e}n}\right|^2\right)\mathcal{F}_{n,+}^{\rm(RR)}\nonumber\\
    && -2\imath{\rm Im}\left[\alpha_{\rm R}^*\left(f_{{\rm o}n}g_{{\rm o}n}^*+f_{{\rm e}n}g_{{\rm
    e}n}^*\right)\mathcal{F}_{n,-}^{\rm(LR)}\right] \bigg\}\ \label{Seh}
\end{eqnarray}
are related to the volume integrals of $|\mathbf{H}_1|^2$ and $(\mathbf{E}_1^*\cdot\mathbf{H}_1)$, respectively.
In Eq.~(\ref{Seh}), we have considered the fact that $({\mu_0/\epsilon_0})^{1/2}=\imath\widetilde{m}\alpha_{\rm R}$.
Equations~(\ref{We})--(\ref{Weh}) are the main analytical results of this study.
Together with the constitutive relations and parameters associated with a single-resonance CMM, Eqs.~(\ref{const1})--(\ref{kappa}) and Eqs.~(\ref{epsilon-eff})--(\ref{kappa-eff}), and the transformation relations from this parameters system to the DBF approach, Eqs.~(\ref{epsilon-dbf})--(\ref{alpha-dbf}), we can readily calculate the average EM energy $\langle W\rangle_t$ within a CMM sphere.
In particular, one can obtain the time-averaged power loss $\langle P\rangle_t$ from the internal energy $\langle W\rangle_t$ by means of the following substitutions in Eqs.~(\ref{We})--(\ref{Weh})~\cite{luan2011}:
\begin{eqnarray}
    \frac{\epsilon_{\rm eff}}{2}&\ \to&\ \omega\epsilon_1''(\omega)= \epsilon_0\left[\frac{\Gamma\omega^2\omega_{\rm p}^2}{\left(\omega_0^2-\omega^2\right)^2+\Gamma^2\omega^2}\right]\
    ,\label{sub1}\\
    \frac{\mu_{\rm eff}}{2}&\ \to&\ \omega\mu_1''(\omega)= \mu_0\left[\frac{\Gamma
    F\omega^4}{\left(\omega_0^2-\omega^2\right)^2+\Gamma^2\omega^2}\right]\
    ,\\
    \frac{\kappa_{\rm eff}}{2}&\ \to&\ \imath\omega\kappa''(\omega)=\frac{-\imath\Gamma
    A\omega^3}{\left(\omega_0^2-\omega^2\right)^2+\Gamma^2\omega^2}\
    ,\label{sub3}
\end{eqnarray}
where $(\epsilon_1'',\mu_1'',\kappa'')={\rm Im}(\epsilon_1,\mu_1,\kappa)$, {\it i.e.}, the imaginary parts of the constitutive quantities.
Also, the scattering, extinction and absorption efficiencies are calculated, respectively, as follows \cite{bohren}:
\begin{eqnarray}
    Q_{{\rm
    sca},\pm}&=&\frac{2}{x^2}\sum_{n=1}^{\infty}(2n+1)\big\{|a_n|^2+|b_n|^2+2|c_n|^2\nonumber\\
    &&\ \pm2{\rm
    Im}\left[\left(a_n+b_n\right)c_n^*\right]\big\}\ ,\label{Qsca}\\
    Q_{{\rm
    ext},\pm}&=&\frac{2}{x^2}\sum_{n=1}^{\infty}\left(2n+1\right){\rm
    Re}\left(a_n+b_n\pm2\imath c_n\right)\ ,\\
    Q_{{\rm abs},\pm}&=&Q_{{\rm ext},\pm}-Q_{{\rm sca},\pm}\ ,
\end{eqnarray}
where the signal ``$+$'' is chosen for RCP waves and ``$-$'' for LCP ones, and the scattering Lorenz-Mie type coefficients $a_n$, $b_n$, $c_n$ and $d_n$ are:
\begin{eqnarray}
a_n&=&\frac{V_n^{\rm(R)}A_n^{\rm(L)}+V_n^{\rm(L)}A_n^{\rm(R)}}{W_n^{\rm(L)}V_n^{\rm(R)}+W_n^{\rm(R)}V_n^{\rm(L)}}\;,\label{an-quiral}\\
    b_n&=&\frac{W_n^{\rm(R)}B_n^{\rm(L)}+W_n^{\rm(L)}B_n^{\rm(R)}}{W_n^{\rm(L)}V_n^{\rm(R)}+W_n^{\rm(R)}V_n^{\rm(L)}}\;,\\
    c_n&=&\imath\frac{W_n^{\rm(R)}A_n^{\rm(L)}-W_n^{\rm(L)}A_n^{\rm(R)}}{W_n^{\rm(L)}V_n^{\rm(R)}+W_n^{\rm(R)}V_n^{\rm(L)}}\;,\\
    d_n&=&-c_n\;,\\
    A_n^{(q)}&=&\widetilde{m}\psi_n(m_{
    q}x)\psi_n'(x)-\psi_n(x)\psi_n'(m_qx)\ ,\\
    B_n^{(q)}&=&\psi_n(m_{
    q}x)\psi_n'(x)-\widetilde{m}\psi_n(x)\psi_n'(m_qx)\ .\label{dn-quiral}
\end{eqnarray}
The mean absorption efficiency $Q_{\rm abs}$ due to both RCP and LCP waves is given by $Q_{\rm abs}=(Q_{\rm abs,L}+Q_{\rm abs,R})/2$.
Following~\cite{ruppin-energy,tiago-joa}, the power loss $\langle P\rangle_t$ is proportional to $Q_{\rm abs}$, and, thereby, depends on $S_{E}$, $S_{H}$ and $S_{EH}$ calculated above.
We {have analytically obtained} that the exact relation between the internal fields and the absorption efficiency is
\begin{eqnarray}
    Q_{\rm abs}&=&\frac{8x}{3}\left[\frac{\epsilon_1''}{\epsilon_{\rm eff}}\frac{\langle W_{E}\rangle_t}{\langle W_0\rangle_t}+\frac{\mu_1''}{\mu_{\rm eff}}\frac{\langle W_{H}\rangle_t}{\langle W_0\rangle_t}+\frac{3}{2}{\rm Re}\left(\widetilde{m}\kappa''\mathcal{S}_{EH}\right)\right]\nonumber\\
    &=&2x\left[\frac{\epsilon_1''}{\epsilon_0}\mathcal{S}_{E}+ \left|\widetilde{m}\right|^2\frac{\mu_1''}{\mu_0}\mathcal{S}_{H}+2{\rm Re}\left(\widetilde{m}\kappa''\mathcal{S}_{EH}\right)\right]\ ,
\end{eqnarray}
where $x=ka$ is the size parameter.
For weakly absorbing spheres with positive refractive index $m=m'+\imath m''$ ($m''\ll m'$), it is well-known that $\langle W\rangle_t /\langle W_0\rangle_t\approx (8/3)m' x (Q_{\rm abs}/m'')$~\cite{bott,tiago-sphere}.
Therefore, in the weak absorption regime, one can approximately calculate the behavior of the internal energy $\langle W\rangle_t$ from $Q_{\rm abs}$.

\section{Numerical calculations}
\label{numerical}

Here we present numerical calculations obtained from a computer code written for the free software {\it Scilab}~5.3.3.
According to Zhao {\it et al.}~\cite{zhao}, we choose realistic parameters for the CMM sphere in the frequency range of Terahertz: $\omega_0=\omega_p=2$~THz and $\Gamma=0.05\omega_0$.
The validity of the effective constitutive quantities $(\epsilon_1,\mu_1,\kappa)$ is guaranteed for size parameters $x=ka\leq1$.
For $\omega\sim10^{12}$~Hz, $ka<1$ is satisfied for $a\sim10^{-4}$~m.
\begin{figure}[htbp]
\centerline{\includegraphics[width=.8\columnwidth]{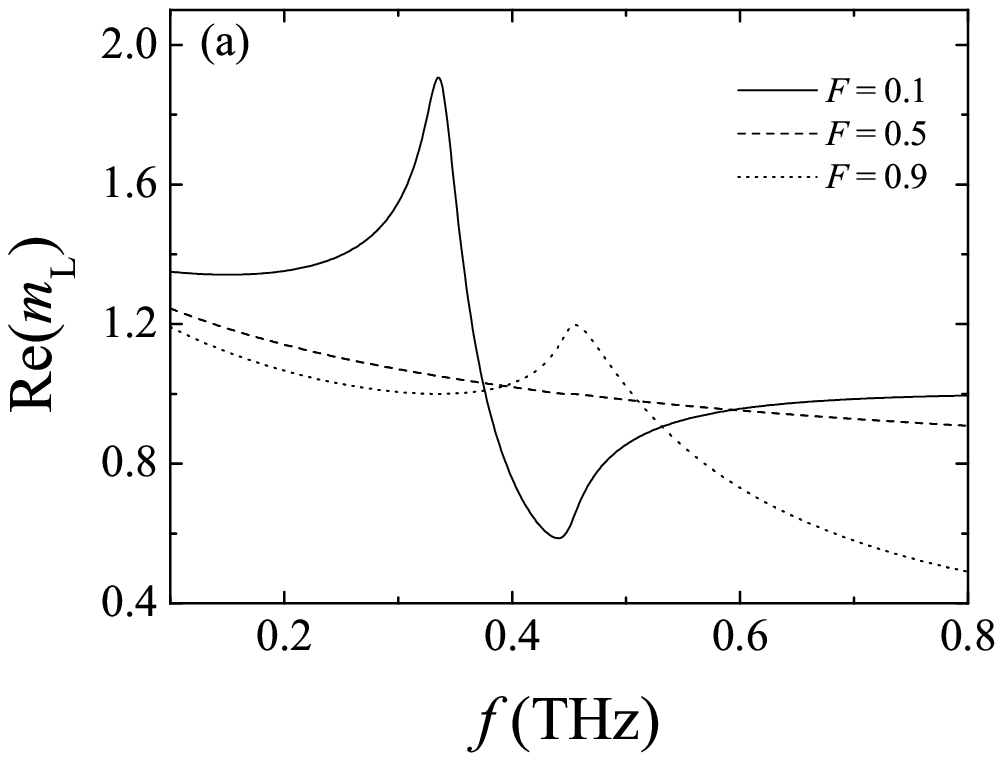}}
\centerline{\includegraphics[width=.8\columnwidth]{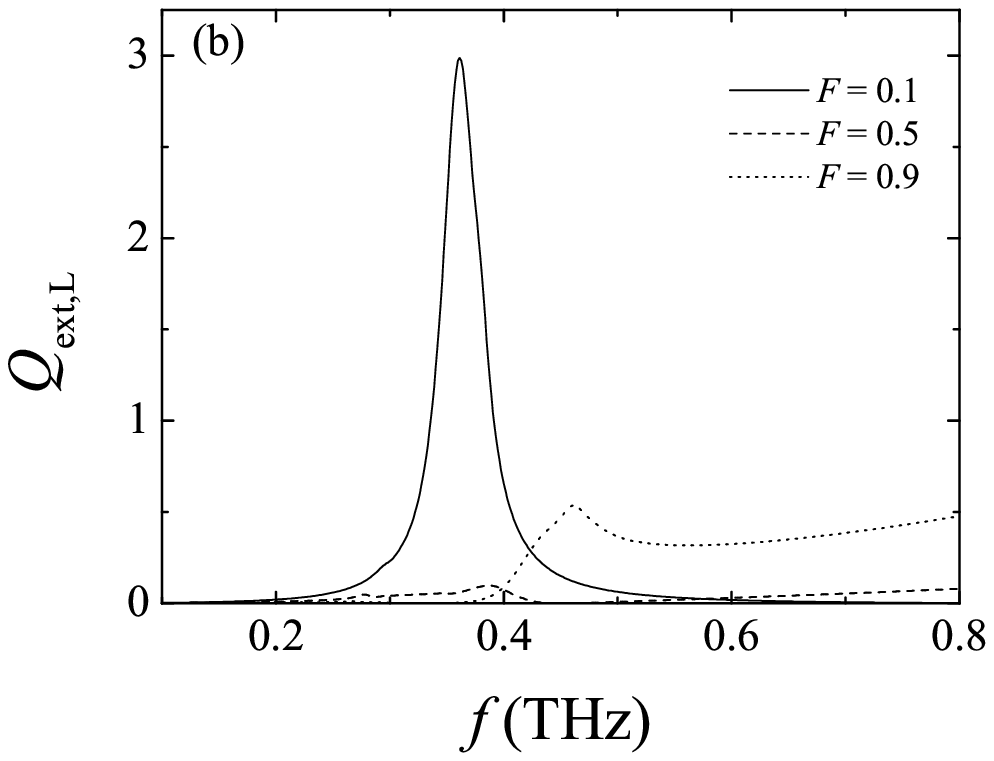}}
\caption{EM scattering by a CMM sphere with radius $a=10^{-4}$~m and $f_0=\omega_0/(2\pi)\approx0.32$~THz $(\omega_{\rm p}=\omega_0)$ as a function of the frequency.
(a) The real part of the LCP refractive index $m_{\rm L}$ for three chirality parameters $\kappa(\omega)$ with filling factors $F=0.1$, 0.5, 0.9.
(b) The corresponding LCP extinction efficiency $Q_{\rm ext, L}$.
There is no negative refractive indices for these parameters.}\label{fig1}
\end{figure}

\begin{figure}[htbp]
\centerline{\includegraphics[width=.8\columnwidth]{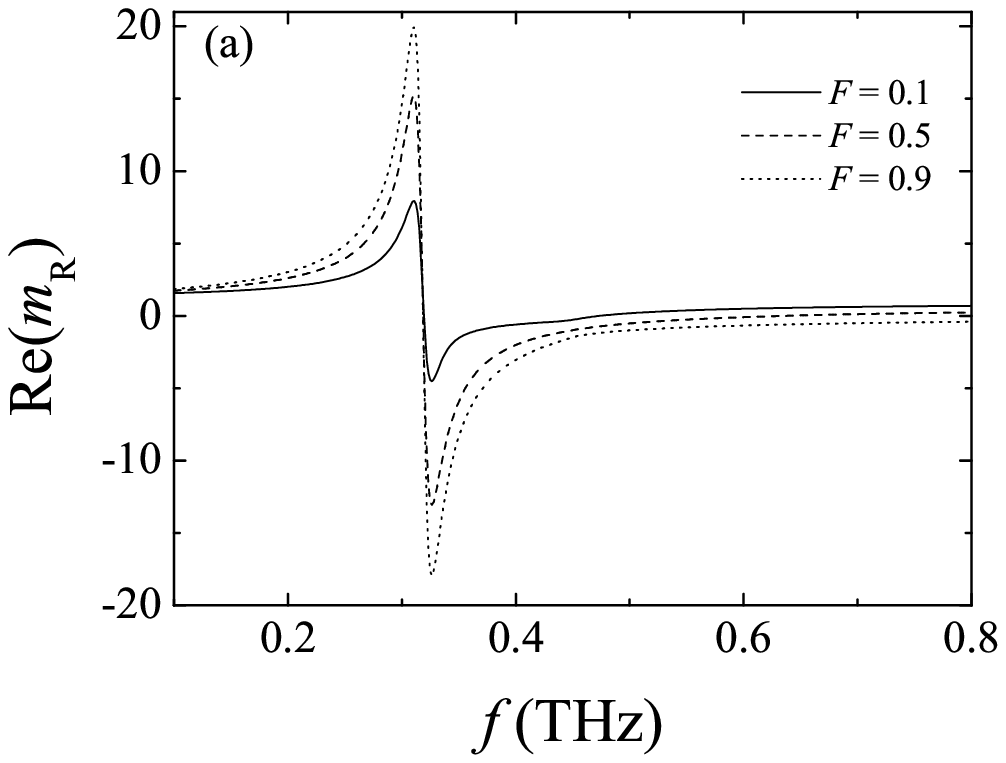}}
\centerline{\includegraphics[width=.8\columnwidth]{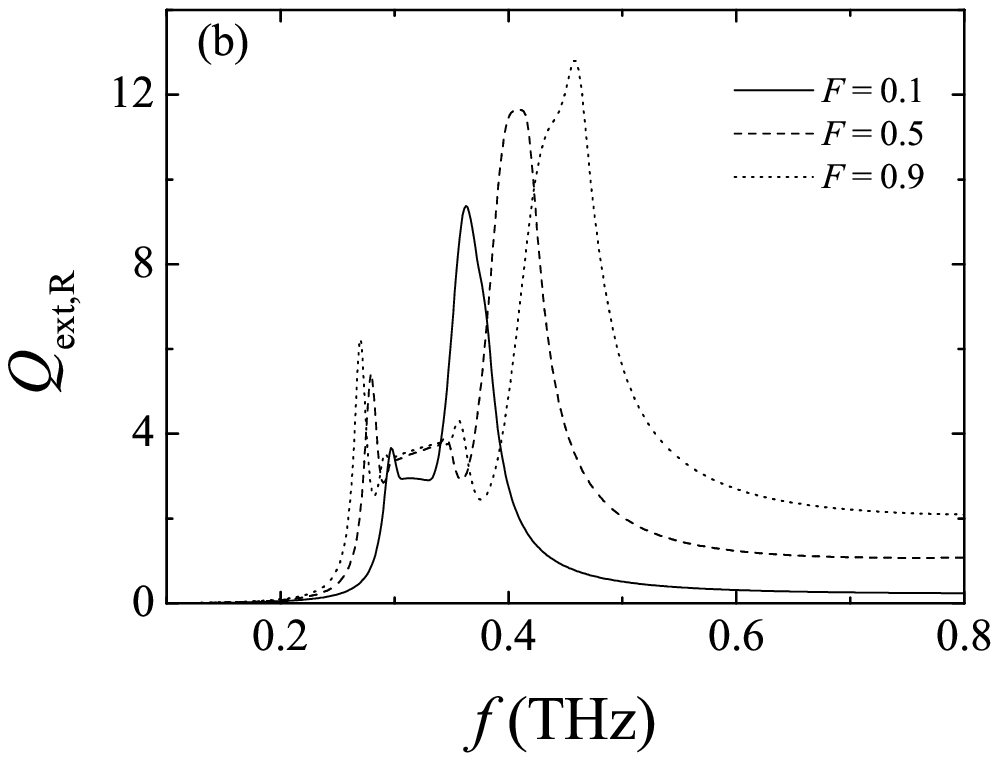}}
\caption{EM scattering by a CMM sphere with radius $a=10^{-4}$~m and $f_0=\omega_0/(2\pi)\approx0.32$~THz as a function of the frequency.
(a) The real part of the RCP refractive index $m_{\rm R}$ for three chirality parameters $\kappa(\omega)$ with filling factors $F=0.1$, 0.5, 0.9.
(b) The corresponding RCP extinction efficiency $Q_{\rm ext, R}$.
The negative refractive indices for RCP waves occur for $f>f_0$.}\label{fig2}
\end{figure}

In Figs.~\ref{fig1} and \ref{fig2}, we show the extinction efficiencies and the associated refractive indices for LCP and RCP waves, respectively, corresponding to EM scattering by a CMM sphere with effective parameters given by Eqs~(\ref{epsilon})--(\ref{kappa}).
Three values of the filling parameter $F$ are considered here: $F=0.1$, 0.5, 0.9.
Note that the real part of $m_{\rm L}$ in Fig.~\ref{fig1}{(a)} does not assume negative values in this frequency range and decreases with the filling parameter $F$ for $f<f_0\approx 0.32$~THz, and increases for frequencies right above $f_0$ ($f=0.4$~THz to $0.5$~THz).
As it can be noted in Fig.~\ref{fig1}{(b)}, $Q_{\rm ext, L}$ decreases with increasing the chirality parameter $|\kappa|\propto{F}^{1/2}$.
The opposite is observed in Fig.~\ref{fig2}.
For frequencies above $f_0\approx 3.2$~THz, the real part of $m_{\rm R}$ is negative and $|m_{\rm R}|$ increases with the chirality parameter $\kappa$, Fig.~\ref{fig2}{(a)}.
This leads to an enhancement in $Q_{\rm ext, R}$, Fig.~\ref{fig2}{(b)}, where the resonance peak for ${\rm Re}(m_{\rm R})<0$ is displaced to high frequencies.

\begin{figure}[htbp]
\centerline{\includegraphics[width=.8\columnwidth]{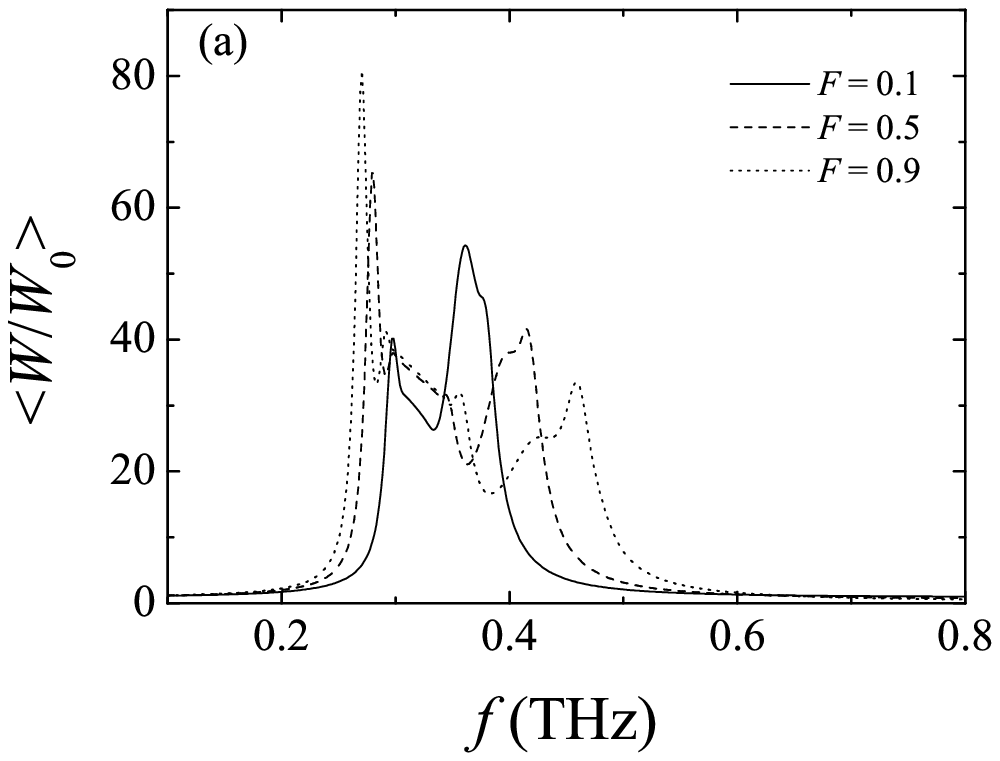}}
\centerline{\includegraphics[width=.8\columnwidth]{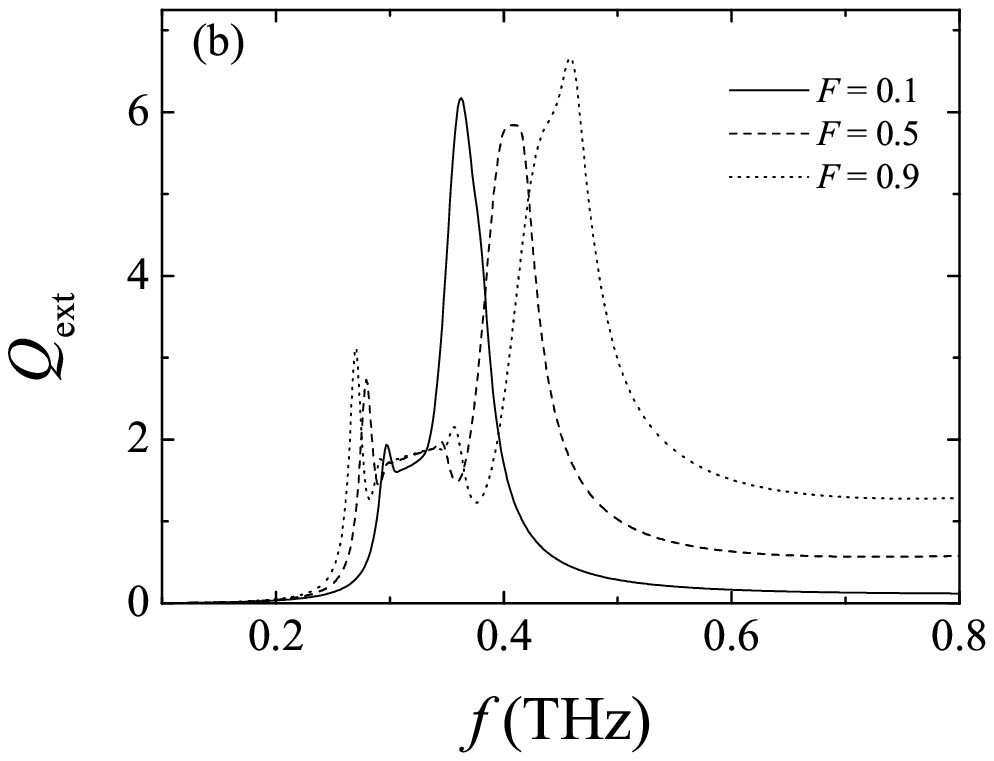}}
\caption{EM scattering by a CMM sphere with radius $a=10^{-4}$~m and $f_0=\omega_0/(2\pi)\approx0.32$~THz as a function of the frequency.
(a) Average energy inside the CMM sphere for filling parameters $F=0.1$, 0.5, 0.9.
(b) The corresponding mean extinction efficiency $Q_{\rm ext}$.
The decrease in $\langle W\rangle_t$ for $f>f_0$ coincides with the increase of $Q_{\rm ext}$. }\label{fig3}
\end{figure}

An interesting result is obtained when we calculate the internal energy, Fig.~\ref{fig3}{(a)}.
Note that, although the mean extinction efficiency $Q_{\rm ext}=(Q_{\rm ext,L}+Q_{\rm ext,R})/2$ in Fig.~\ref{fig3}{(b)} is increasing with the chirality parameter $\kappa$ for $f>f_0$ and ${\rm Re}(m_{\rm R})<0$, the internal energy is decreasing simultaneously.
For a sufficiently strong chirality parameter, the LCP waves tend to be suppressed, as can be seen in Fig.~\ref{fig1}{(a)}.
Correspondingly, strong chirality favors the reflection of RCP waves on the interface between the CMM sphere and the surrounding medium.
Since $Q_{\rm ext}$ increases as $\kappa$ increases for ${\rm Re}(m_{\rm R})<0$, we conclude that incident waves are strongly reflected by the surface of the sphere and, for this reason, do not contribute to the internal energy.

\begin{figure}[htbp]
\centerline{\includegraphics[width=.8\columnwidth]{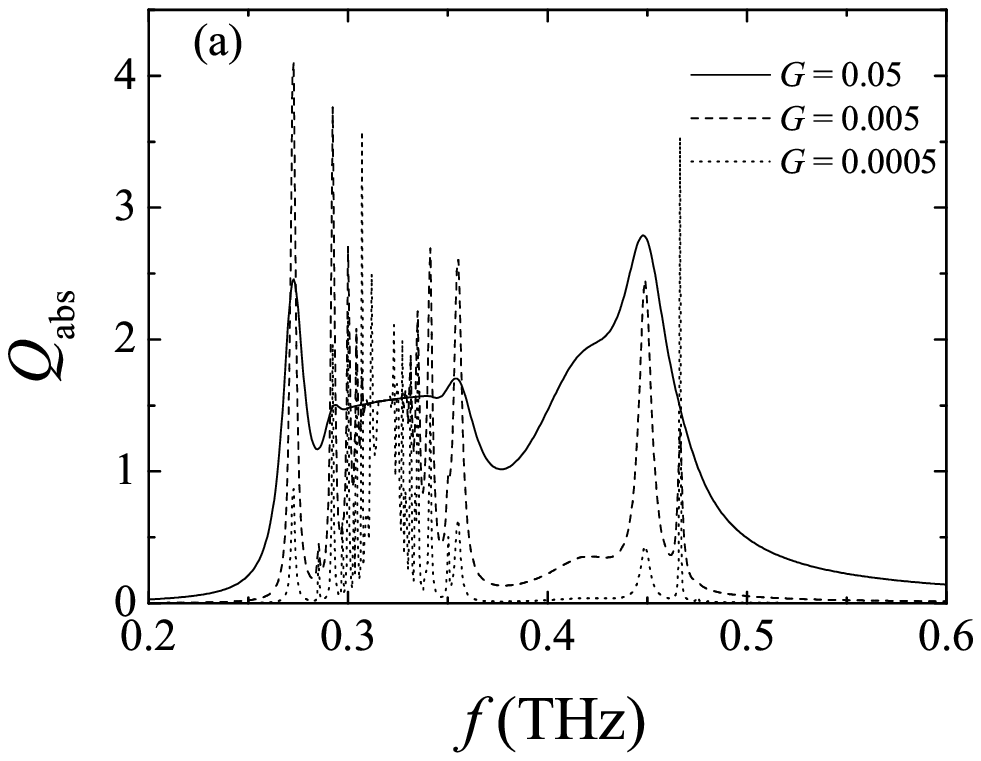}}
\centerline{\includegraphics[width=.8\columnwidth]{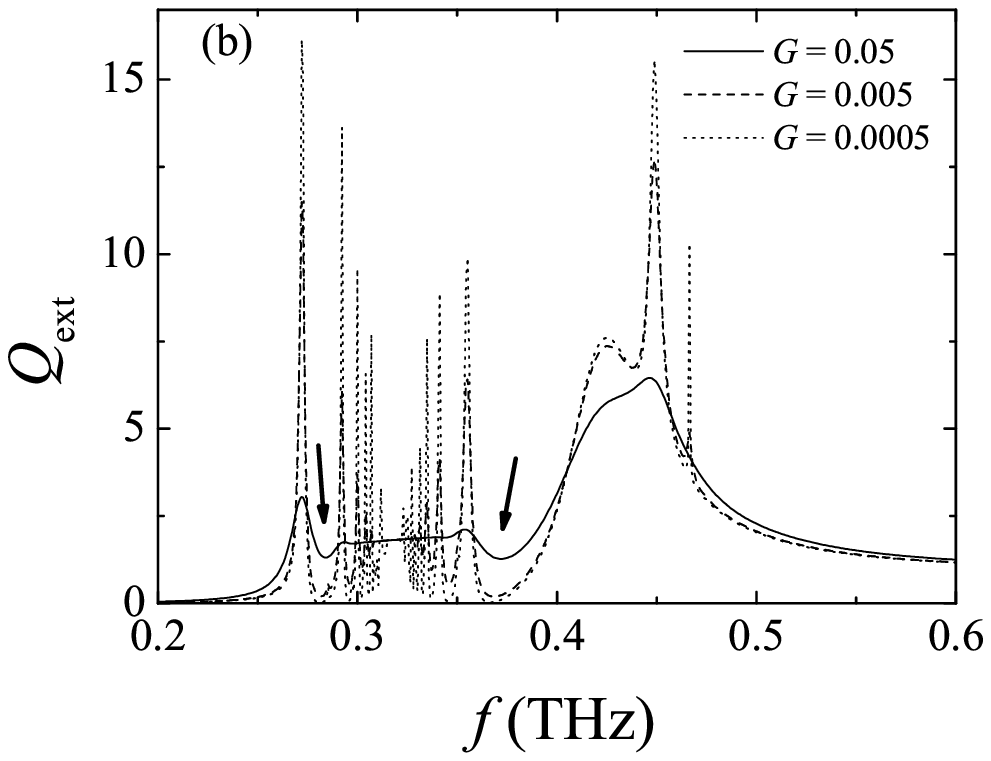}}
\centerline{\includegraphics[width=.8\columnwidth]{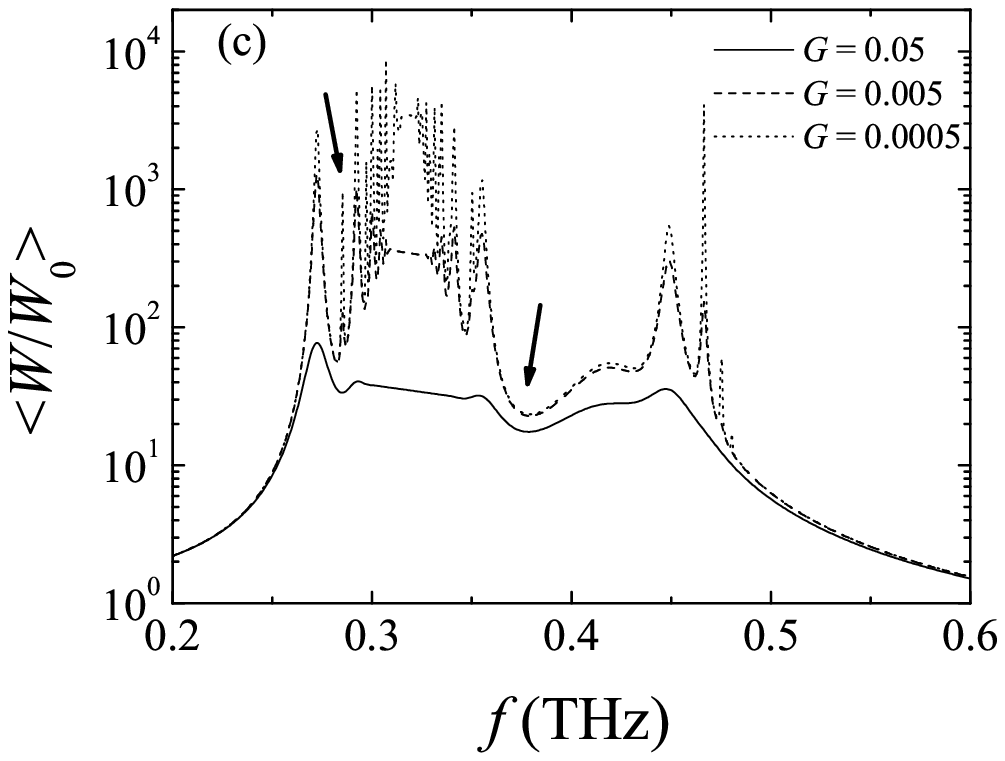}}
\caption{CMM sphere with radius $a=10^{-4}$~m, filling factor $F=0.8$ and $\omega_0=\omega_{\rm p}=2$~THz as a function of frequency and the absorption parameter $G=0.05$, 0.005, 0.0005.
(a) The mean absorption efficiency $Q_{\rm abs}$.
(b) The mean extinction efficiency $Q_{\rm ext}$.
(c) The internal energy $\langle W\rangle_t$.
For $f>f_0\approx0.32$~THz, a decrease in $Q_{\rm abs}$ and $Q_{\rm ext}$ do not necessarily coincide with a decrease in $\langle W\rangle_t$.}\label{fig4}
\end{figure}

In Fig.~\ref{fig4}, we use the same parameters as above, assuming now $F=0.8$ (``strong'' chirality) as a function of the absorption coefficient $\Gamma=G\omega_0$, where $G=0.05$, 0.005, 0.0005.
Here, we notice some regions where weak absorption ($Q_{\rm abs}\approx0$) leads to weak extinction ($Q_{\rm ext}\approx0$), and increases the amount of energy stored in the CMM sphere.
This effect occurs above $f_0$, in the range $0.37$~THz$<f<0.40$~THz, see Figs.~\ref{fig4}{(a)} and \ref{fig4}{(b)}.
For $f\approx0.37$~THz, $Q_{\rm abs}$ and $Q_{\rm ext}$ almost vanish for $G=0.0005$, but the value of $\langle W\rangle_t$ in Fig.~\ref{fig4}{(c)} is still large.
In addition, there are resonance peaks in the internal energy around $f_0$ that do not correspond to a peak in $Q_{\rm ext}$, such as the one at $f\approx0.28$~THz.
This behavior is indicated by a solid arrow in Figs.~\ref{fig4}{(b)} and \ref{fig4}{(c)} and {suggests} the existence of off-resonance field enhancement within the sphere.
These resonances are related to interferences between two different Lorenz-Mie type coefficients, such as $f_ng_n^*$ in $\langle W\rangle_t$, Eq.~(\ref{S}), and $a_nc_n^*$ in $Q_{\rm sca}$, Eq.~(\ref{Qsca}), and are explained in terms of the Fano effect~\cite{luk}.
This effect also occurs for coated spheres containing dispersive Drude-like materials with low absorption~\cite{miroshnichenko}.

\section{Conclusions}
\label{conclusions}

We have calculated the time-averaged energy inside a CMM sphere using a set of dispersive realistic parameters provided by an effective medium theory~\cite{zhao}.
Using the decomposition of the EM field in LCP and RCP waves~\cite{bohren-active-sphere}, we have obtained the internal fields for the magnetic case and the respective Lorenz-Mie coefficients in DBF approach.
Since the energy density in an isotropic chiral media is commonly calculated from the Condon constitutive relations~\cite{condon}, a transformation of parameters has been applied to obtain the exact expression for the internal energy. We have expressed the absorption efficiency in terms of the internal energy and the constitutive parameters. Although the internal energy decreases with the chirality parameter in the region of negative refraction for RCP waves, the mean extinction efficiency increases as the chirality parameter increases.
Our results reveal that, in the small-particle limit, off-resonance field enhancement occurs for weakly absorbing spheres with strong chirality. We show that this effect is due to field interferences between different multipolar moments, which lead to Fano resonances in electromagnetic scattering. We hope that our findings could be exploited in applications involving dispersive CMM scatterers and Fanoshells.

\section*{Acknowledgments}

The authors acknowledge the Brazilian agencies for support.
TJA holds grants from Funda\c{c}\~ao de Amparo \`a Pesquisa do Estado de S\~ao Paulo (FAPESP) (2010/10052-0), ASM from Conselho Nacional de Desenvolvimento Cient\'{\i}fico e Tecnol\'ogico (CNPq) (305738/2010-0), and FAP from Funda\c{c}\~ao de Amparo \`a Pesquisa do Estado do Rio de Janeiro (FAPERJ) (E-26/111.463/2011).

\end{document}